\documentclass[showpacs,amsmath,amssymb,showkeys,pre]{revtex4}

\newcommand{\beq}{\begin{equation}}
\newcommand{\eeq}{\end{equation}}
\newcommand{\beqa}{\begin{eqnarray}}
\newcommand{\eeqa}{\end{eqnarray}}
\newcommand{\non}{\nonumber}

\begin{document}

\title{The role of the Becchi-Rouet-Stora-Tyutin supersymmetry in the calculation of the
complexity for the Sherrington-Kirkpatrick model}

\author{Alessia Annibale, Andrea Cavagna, Irene Giardina, Giorgio Parisi, Elisa Trevigne}

\affiliation{Dipartimento di Fisica, Universit\`a di Roma ``La Sapienza'' and \\
Center for Statistical Mechanics and Complexity, INFM Roma 1, \\
Piazzale Aldo Moro 2, 00185 Roma,  Italy}

\date{July 18, 2003}

\begin{abstract}
The Becchi-Rouet-Stora-Tyutin (BRST) supersymmetry is a powerful tool for the calculation 
of the complexity of metastable states in glassy systems, and it is particularly useful to
uncover the relationships between complexity and standard thermodynamics. In this work
we compute the Thouless-Anderson-Palmer complexity of the Sherrington-Kirkpatrick 
model at the quenched level, by using the BRST supersymmetry.
We show that the complexity calculated at $K$ steps of replica symmetry breaking is
strictly related to the static free energy at $K+1$ steps of replica symmetry breaking.
The supersymmetry therefore provides a prescription to obtain the complexity of the TAP
states from the standard static free energy, even in models which are solved by more 
than one step of replica symmetry breaking. This recipe states that the complexity is 
given by the Legendre transform of the static free energy, where the Legendre parameter is 
the largest replica symmetry breaking point of the overlap matrix.
\end{abstract}

\pacs{05.50.+q,  75.10.Nr,  12.60.Jv }
\keywords{Spin-Glasses, TAP Equations, Supersymmetry}

\maketitle

\section{Introduction}

Glassy systems in their low temperature phase always display a very complicated structure
of metastable states. In fact, this feature can be adopted more generally as the very 
definition of a {\it complex system}, that is a system with a highly nontrivial structure
of its energy, or free energy landscape. Of course, this definition is not limited to 
systems with an energy or free energy function, but can be given in general for any system
described by a global state function, whatever this function is: a fitness function in 
biology, a cost function in optimization problems, and so on. 
A common feature to all complex systems, is that the number of metastable states is
exponentially large in the size of the system, in such a way that it is possible to define
an entropy density $\Sigma$ of the metastable states, normally called {\it complexity}.
The calculation of the complexity is in general a crucial task for understanding
both the static and dynamical features of complex systems.

In the context of spin-glasses, the problem of calculating the complexity is quite an old 
one \cite{bm1,bm2,bm3,dd1,innocent,ddy,bmy}. 
In particular, starting with the classic paper by Bray and Moore \cite{bm1}, a large number 
of investigations have focused on the metastable states of the Thouless-Anderson-Palmer 
(TAP) free energy \cite{tap} in the Sherrington-Kirkpatrick (SK) model \cite{sk}. 
Despite all these studies, there are still some open questions regarding the TAP complexity 
in the SK model, mostly related to the consistency of the TAP approach with the standard 
static calculation of the thermodynamic free energy. A first question, addressed by 
De Dominicis and Young in \cite{ddy}, is whether the partition function computed within 
the TAP approach coincides with the standard thermodynamic one. The result of such a study 
was that the two partition functions are the same only if one imposes some identities, whose 
origin was unclear at the time. On the other hand, in \cite{bm2} and \cite{bmy} the formal
relationships between the calculation of the TAP complexity and that of the static free
energy were investigated. These studies clearly pointed out that these two quantities
were closely related, but were unable to establish an exact formal connection between them.
In particular, the equations involved in the calculation of the complexity are practically
intractable in the SK model, where a full replica symmetry breaking solution must be adopted.

In a recent work, the problem of the SK complexity has been reconsidered by using a supersymmetric
approach \cite{brst1}. More precisely, it was noted in \cite{juanpe} that the effective action
used to compute the number of TAP states with fixed free energy density $f$ is invariant under
a generalization of the BRST \cite{brs, tito,zj} supersymmetry. This invariance can be used to generate 
a set of Ward identities which significantly reduces the multiplicity of solutions of the saddle
point equations, thus simplifying considerably the calculation of the complexity $\Sigma$.
The first important result of \cite{brst1} was to show that the identities used by De Dominicis 
and Young to prove the consistency of TAP and standard partition functions, were in fact the 
BRST Ward identities. Moreover, this supersymmetric approach has been used in \cite{brst1} 
to compute the {\it annealed} complexity of the SK model, which turned out to be exactly 
connected to the quenched static free energy at the one step of replica symmetry breaking ($1$-RSB).
This result suggested the existence of a deep connection between TAP complexity 
and static free energy, irrespective of the degree of approximation we use to compute them. 

What we show in the present work, is that indeed such a connection is present. 
The {\it quenched} supersymmetric calculation of the TAP complexity in the SK model 
turns out to be completely equivalent to
the calculation of the standard free energy, with some peculiar connections between the replica
symmetry breaking structure of the overlap matrices in the two approaches. This proves at the
deepest level the equivalence of the TAP and static approaches in the SK model, suggesting that 
such an equivalence may be valid in any glassy system, at least at the mean-field level.

There is a further motivation to reconsider the connections between complexity and static free
energy. In \cite{monasson,franzparisi} a novel method to compute the complexity was introduced, 
which is
independent from the TAP approach, and thus is more viable to be used in system where a 
mean-field TAP free energy cannot be defined. Of course, a key issue is whether this 
alternative definition of the complexity is in general equal to the TAP complexity in those
systems where a TAP free energy exists. According to this alternative method, in the 
formulation of \cite{monasson}, the complexity is given 
by the Legendre transform of the static free energy of $r$ systems forced to be in the same state.
This unusual free energy has been put in connection in the past with some ``replicated'' versions
of the standard static free energy \cite{potters,potters2} and this Legendre complexity has been 
thus computed and 
compared with the TAP complexity. The results have been immediately clear in $1$-RSB systems, where
the two complexities clearly coincide \cite{crisatap}. However, this issue was up to now 
 far less clear in full-RSB systems, 
as the SK model: due to the nontrivial form of static the overlap matrix, it is not obvious
what is the correct form of the replicated free energy that has to be used to 
compute the complexity by means of the Legendre transform method.

The study we perform in this work shows that supersymmetric TAP complexity and Legendre complexity 
coincide in the SK model at any level of replica symmetry breaking. More specifically,
by using the BRST supersymmetry we show analytically that in order to obtain the complexity 
at the $k$ RSB level, we have to perform the Legendre transform of a standard static free
energy calculated at the $k+1$ RSB level. The Legendre parameter is the $k+1$-th symmetry
breaking point, that is the largest breaking point of the static overlap matrix $Q_{ab}$.
This allows a calculation of the complexity starting from the full-RSB form of the static
free energy of the SK model. Work in this direction is in progress \cite{leuzzi-full}.

In section 2 we briefly review the two methods for computing the complexity, and show how, in 
the TAP context, they lead to the same result. Both methods formally require the calculation 
of the free energy of $r$ systems forced to be in the same TAP state, and this is thus the
quantity we calculate in section 3 by using the BRST supersymmetry. In section 4 we finally
show how the quantity we have obtained is related to the standard free energy of the SK model,
thus giving a general prescription to compute the complexity starting from the static free energy
within the Legendre transform method. 
In section 4.B can be found the main original result of our work, namely the formal connection
between complexity at $k$ RSB level and static free energy at $k+1$ RSB level.
Conclusions are discussed in section 5. A shorter account of our results can be found in \cite{corto}.

\section{Different methods, same complexity}

In this section we will briefly review the two different methods to compute the complexity, 
and discuss their mutual connections. 
The first method \cite{bm1} is only defined when a mean-field free 
energy, function of the local magnetizations $m_i$, is defined. This quantity is known in the spin-glass 
context as TAP free energy $F_{TAP}$, and its local minima $\{m_i^\alpha\}$, 
labeled by $\alpha=1,\dots,\cal N$ 
are identified with the metastable states of the system. 
The complexity $\Sigma(\beta, f)$ of the TAP states 
with free energy density $f$, at inverse temperature $\beta$, is defined as,
\beq
\Sigma(\beta,f) 
= \frac{1}{N}\log \sum_{\alpha=1}^{\cal N} \delta[\beta N f - \beta F_{TAP}(m_\alpha)] 
= \frac{1}{N}\log \int du\ e^{N u \beta f} \sum_{\alpha=1}^{\cal N} e^{-u \beta F_{TAP}(m_\alpha)} \ ,
\eeq
where $\alpha$ indicates a given metastable TAP state. 
If we compare this equation with the equivalent one in the past investigations of \cite{bm1,bm2,bm3,bmy,brst1},
we can see that we have made the change of notation $u\to -u$. The technical reason for this will be clear
later. We hope that this change of notation will not make difficult the comparison with 
former studies. 
If we define the thermodynamic potential $\Phi(\beta,u)$
as,
\beq
\exp\left( -\beta N u \, \Phi \right)
\equiv
\sum_{\alpha=1}^{\cal N} e^{-\beta u\, F_{TAP}(m_\alpha)} \ ,
\label{phidef}
\eeq
we can use the steepest descent method to obtain,
\beq
\Sigma(\beta,f) = \beta u f -\beta u\,\Psi(\beta,r) \ , 
\label{heidi}
\eeq
where the parameter $u=u(\beta,f)$ is fixed by the equation,
\beq
\Phi(\beta,u) + u \,\frac{\partial\Phi(\beta,u)}{\partial u}=f \ .
\label{nonno}
\eeq
In other words, the TAP complexity can be obtained as the Legendre transform of the effective 
thermodynamic potential $\Phi(\beta,u)$ with respect to the parameter $u$, which therefore 
is the Legendre-conjugate variable of the free energy density $f$.

A different approach to the calculation of the complexity, which does not
a priori rely on the existence of a TAP free energy, is the one introduced in \cite{monasson}.
The total equilibrium free energy of a super-system composed by $r$ real replicas forced to stay in 
the same state is given by,
\beq
r\,\Psi(\beta,r) \equiv -\frac{1}{\beta N}\log\int df \ e^{N\Sigma(\beta,f)} \, e^{-\beta N r f} = 
{\rm Ext}_f \, \left[ rf - T \Sigma(\beta,f )\right] \ .
\label{remi}
\eeq
Note that $\Psi(\beta, r)$ is the free energy density {\it per replica} of such a super-system. 
The key idea of this approach is that, due to the constraint to stay in the same state, there is no degeneracy 
of the complexity term in the previous formula, and thus by tuning $r$, we can tune the saddle point over $f$,
spanning the entire free energy spectrum of metastable states. More precisely, this amounts to say that we can invert 
relation (\ref{remi}) and find the complexity as the Legendre transform with respect 
to the parameter $r$ of the thermodynamic potential $\Psi(\beta,r)$, 
\beq
\Sigma(\beta, f)=  \beta r f - \beta r\, \Psi(\beta,r) \ ,
\label{pollon}
\eeq
with,
\beq
\Psi(\beta,r) + r \,\frac{\partial\Psi(\beta,r)}{\partial r}=f \ .
\label{nonnino}
\eeq
Although the formal similarity between equations (\ref{heidi}) and (\ref{pollon}) is clear, it may not be as
clear what are the {\it physical} connections between the thermodynamic potentials $\Phi(\beta,u)$ and 
$\Psi(\beta,r)$.
In particular, in a standard static approach it is not obvious how to compute the constrained free energy 
$\Psi$. A possibility is to couple in some way the $r$ real replicas in the Hamiltonian, and then let the
coupling go to zero. As an alternative, we can simply compute the {\it normal} free 
energy of a set of $r$ systems by computing the average replicated partition function, $\overline{Z^{r n}}$, 
and then impose the constraint on the $r$ replicas at the level of the overlap matrix \cite{potters,potters2}. 
Of course, this needs some careful breaking of the replica symmetry, otherwise we simply have,
\beq
\lim_{n\to 0}\frac{1}{n}\log\overline{Z^{r n}}=\lim_{n\to 0}\frac{1}{n}\log[1+ r\,n\ \overline{\log Z}]=
-\beta r F \neq -\beta r \Psi(r) \ .
\eeq
On the other hand, in the TAP context it is straightforward to force the $r$ replicas to be 
in the same state. This can simply be done by computing a partition function where we sum over
all the $r$ configurations belonging to the same TAP state $\alpha$, and then sum over just {\it one} 
set of TAP states. Such a partition function can be written as,
\beq
Z(r)=
\sum_{\alpha=1}^{\cal N}\ \sum_{\sigma_1\in\alpha} e^{-\beta H(\sigma_1)} \dots 
\sum_{\sigma_r\in\alpha} e^{-\beta H(\sigma_r)} \ .
\eeq
Note that of course $Z(r)\neq Z^r$, unless some nontrivial replica symmetry breaking procedure is adopted.  
The thermodynamic potential $\Psi(\beta, r)$ is therefore defined in the TAP context by the formula,  
\beq
\exp(-\beta N r \,\Psi )  = Z(r)=
\sum_{\alpha=1}^{\cal N} e^{-\beta r F_{TAP}(m_\alpha)} \ ,
\label{psidef}
\eeq
where we have used the standard TAP relation \cite{ddy},
\beq
\sum_{\sigma\in\alpha} e^{-\beta H(\sigma)} = e^{-\beta F_{TAP}(m_\alpha)} \ .
\eeq
By comparing (\ref{phidef}) and (\ref{psidef}) it is now clear that $\Psi(\beta,r)$ and 
$\Phi(\beta,u)$ have exactly the same formal definition, with $u=r$. 
This identification justifies our choice to use $u$ rather than $-u$ as it was done in 
\cite{bm1,bm2,bm3,bmy,brst1}, such that we simply have,
\beq
\Psi(\beta,r)= \Phi(\beta,u) \ .
\label{indo}
\eeq
Therefore, in the TAP context the two definitions of the complexity coincide, and the key quantity to compute
is the thermodynamic potential $\Psi(\beta,r)$ (or, equivalently, $\Phi(\beta,u)$), 
that is the Legendre transform of the complexity with respect 
to the free energy density $f$. This quantity is what we calculate in the 
next section by using the BRST supersymmetry. 
What we will find is that in the SK model $\Psi(\beta,r)$ is deeply connected to 
the standard static free energy of 
the system.

\section{The supersymmetric quenched calculation}

In this section we will calculate the potential $\Psi(\beta,r)$ in the TAP context, by using 
equation (\ref{psidef}). We will follow the general method introduced by Bray and Moore in \cite{bm1},
and we will use the BRST supersymmetry, firstly introduced in the TAP context in \cite{juanpe}, and 
discussed for the SK model in \cite{brst1}. 
We warn once again the reader that for a comparison with the previous calculations
of \cite{bm1,bm2,bm3,bmy,brst1} one has to set $r=-u$, since the potential which was normally calculated 
in the past was $\Psi(\beta,-u)$, rather than $\Psi(\beta,r)$.

\subsection{The calculation}

The TAP free energy for the SK model is given by \cite{tap},
\beq
F_{TAP}(m)=-\frac{1}{2} \sum_{ij} J_{ij} m_i m_j + \frac{1}{\beta} \sum_i \phi_0(q,m_i) \ ,
\label{ftap}
\eeq
with,
\beq
\phi_0(q,m) =
\frac{1}{2}\log(1-m^2) + m\,\tanh^{-1}(m) -\log 2 -\frac{\beta^2}{4}(1-q)^2 \ .
\label{phi0}
\eeq
The variables $m_i$ are the local magnetizations, and $q$ is  the self-overlap of the TAP states,
\beq
q=\frac{1}{N}\sum_i m_i^2 \ ,
\eeq
while the quenched couplings $J$ are random variables with Gaussian distribution,
\[
P(J_{ij})=\sqrt{N/2\pi}\ \exp(-NJ_{ij}^2/2) \ .
\non
\]
The TAP equations and the Hessian of the free energy are respectively,
\beqa
\beta \, \partial_i F_{TAP}(m) &=& -\beta \sum_{j\neq i} J_{ij} m_j + \phi_1(q,m_i) = 0 \ , \non\\
\beta \, \partial_i \partial_j F_{TAP}(m) &=& -\beta J_{ij} + \phi_2(q,m_i) \, \delta_{ij} \non \ ,
\eeqa
with,
\beqa
\phi_1(q,m) &=& \beta^2 (1-q) m + \tanh^{-1}(m) \non \\
\phi_2(q,m) &=& \beta^2 (1-q) + \frac{1}{1-m^2} \ + \,O(1/N) \ .
\label{phi}
\eeqa
The term of order $1/N$ in $\phi_2(q,m)$ will be dropped in what follows. From (\ref{psidef}) we have that 
the {\it quenched} definition of the potential $\Psi(\beta,r)$ is given by,
\beq
- \beta r\, \Psi(\beta,r)=
\frac{1}{N}\, \overline{\log\rho(\beta,r|J)} = \frac{1}{Nn}\log\overline{\rho(\beta,r|J)^n} \ ,
\eeq
with $N\to\infty$ and $n\to 0$, and where, 
\beq
\rho(\beta,r|J)=\sum_{\alpha=1}^{\cal N} e^{-\beta r\,F_{TAP}(m_\alpha)}
=\int \prod_i dm_i\ \delta(\partial_i F_{TAP}(m)) \ |\det (\partial_i \partial_j F_{TAP}(m))| 
\ e^{-\beta r\,F_{TAP}(m)} \ .
\label{rho}
\eeq
We will perform the calculation following closely the lines of \cite{brst1}, and more
generally the methods developed in the past for this kind of calculation \cite{tap, bm1, ddy}.
In particular, the modulus of the Hessian determinant will be dropped. This approximation 
is safe only in the energy/temperature region where minima  are dominant with respect to unstable 
saddles. We do not know to what extent this is true in the SK model, but we expect it to be true close 
to the ground state (the lower band edge). Since we are interested in the consistency of the calculation 
of the complexity with the statics at the lower band edge, dropping the modulus should be reasonably safe.
However, we stress again that this method (as any other method which disregards the modulus) is {\it not}
under control if saddle points of the TAP free energy are exponentially dominant over stable minima: in 
that case we are weighting each stationary point with the undefined sign of its Hessian determinant, 
with results which are hard to forecast. For a deeper discussion of this point see \cite{jorge1}. 
After introducing the commuting (Bosonic) fields $x_i$ to implement the delta function,
and the anti-commuting (Fermionic) fields $\bar\psi_i,\psi_i $ for the determinant, we find,
\beq
\rho(\beta,r|J)
= \int {\cal D}m\, {\cal D}x\,{\cal D}\bar\psi\, {\cal D}\psi\, du \ \
e^{\beta S(m,x,\bar\psi,\psi,r)} \ ,
\label{pino}
\eeq
where the action $S$ is given by,
\beq
S(m,x,\bar\psi,\psi)= \sum_i x_i \partial_i F_{TAP}(m) + \sum_{ij}\bar\psi_i  \psi_j  
\partial_i \partial_j F_{TAP}(m) - r F_{TAP}(m) \ .
\label{action}
\eeq
By averaging $\rho(\beta,r|J)^n$  over the disorder we obtain 
the following effective action,
\beqa
\beta S &=& \frac{\beta^2}{2 N} \left[\sum_{ab}^n \left(\sum_i^N x_i^a x_i^b\right) 
\left(\sum_j^N m_j^a m_j^b\right) +\sum_{ab}^n \left(\sum_i^N x_i^a m_i^b\right)^2 - 
\sum_{ab}^n \left(\sum_i^N \bar\psi_i^a \psi_i^b\right)^2\right] \
\non
\\
&+&  \frac{\beta^2}{2 N} \left[\frac{r^2}{2} \sum_{ab}^n \left(\sum_i^N m_i^a m_i^b\right)^2 -
2 r \sum_{ab}^n \left(\sum_i^N m_i^a x_i^b\right) \left(\sum_j^N m_j^a m_j^b\right)\right] 
\non
\\
&+& \sum_a^n \sum_i^N \left[x_i^a \phi_1(m_i^a) + 
\bar\psi_i^a \psi_i^b \phi_2(m_i^a) - r  \phi_0(m_i^a)\right] 
\eeqa
In order to linearize the quadratic terms we follow the standard method to introduce the 
Lagrange multipliers,
\beqa
\delta(q_{ab} N - \sum_i m_i^a m_i^b) &=&\int_{-i\infty}^{+i\infty} \
\frac{d\lambda^{ab}}{2\pi i} \ e^{-\lambda^{ab} q_{ab} N + \lambda^{ab} \sum_i m_i^a m_i^b} 
\non
\\
\non
\delta(W_{ab} N-\sum_i m_i^a x_i^b)&=&
\int_{-i\infty}^{+i\infty}\frac{dr^{ab}}{2\pi i} \ e^{-w^{ab} W_{ab} N + w^{ab}\sum_i m_i^a x_i^b} \\
\non
\delta(T_{ab} N-\sum_i \bar\psi_i^a\psi_i^b)&=& \
\int_{-i\infty}^{+i\infty} \frac{dt^{ab}}{2\pi i} \ e^{-t^{ab} T_{ab} N + t^{ab}\sum_i \bar\psi_i^a\psi_i^b} \\
\delta(L_{ab} N-\sum_i x_i^a x_i^b) &=& \
\int_{-i\infty}^{+i\infty} \frac{dl^{ab}}{2\pi i} \ e^{-l^{ab} L_{ab} N + l^{ab} x_i^a x_i^b}
\label{deltas}
\eeqa
and we integrate over $q_{ab},W_{ab},T_{ab}$ and $L_{ab}$.
In this way the integrals in $x_i$ and $(\bar\psi_i,\psi_i)$ become
Gaussian and can be performed explicitly, giving,
\beq
\overline{\rho(\beta,u|J)^n}
=\int {\cal D}\Omega \ e^{N\Sigma_0(\Omega) + N \log\int \prod_a dm^a \ 
e^{{\cal L}(\Omega, m^a)}} \ ,
\eeq
where $\Omega=\{q_{ab},\lambda^{ab},w^{ab},W_{ab},t^{ab},T_{ab},L_{ab},$\
$l^{ab}\}$  and,
\beqa
\Sigma_0(\Omega)&=&
\frac{\beta^2}{2} \sum_{ab} \left[q_{ab} L_{ab} + W_{ab}^2 - T_{ab}^2 + \frac{r^2}{2}\,q_{ab}^2
- 2 r\, q_{ab} W_{ab} \right]
\non 
\\
&& - \sum_{ab} \left[ \lambda^{ab} q_{ab} + w^{ab} W_{ab} + t^{ab} T_{ab} + l^{ab} L_{ab}  \right] 
-\frac{1}{2}\log\left[(4\pi)^n\det(l^{ab})\right]
\label{pezzo1}
\\
{\cal L}(\Omega,m^a)&=& -r \sum_a \phi_0(q_{aa},m^a) 
-\frac{1}{4}\sum_{ab} \left[\phi_1(q_{aa},m^a)+\sum_c w^{ac} m^c\right]
l_{ab}^{-1} \left[\phi_1(q_{bb},m^b)+\sum_c w^{bc} m^c\right] \
\non
\\
 &+&\log\left[\det(\phi_2(q_{aa},m^a) \delta_{ab}+t^{ab})\right]+ \
\sum_{ab} \lambda^{ab} m^a m^b  \ .
\label{pezzo2}
\eeqa
Thanks to the prefactor $N$ in the exponential, the integral in 
${\cal D}\Omega$ can be performed with the steepest descent method.
In this way we can write the quenched thermodynamic potential $\Psi(\beta,r)$ as,
\beq
- \beta r\,\Psi(\beta,r) = \lim_{n\to 0} \, \frac{1}{n}\, \left[\, \Sigma_0 (\hat\Omega)+
\log\int\prod_a dm^a e^{{\cal L}(\hat\Omega, m^a)}\, \right] \ ,
\eeq
where, as usual,  $\hat\Omega$ is solution of the saddle point equations,
\beq
0=
\frac{\partial \Sigma_0(\hat\Omega)}{\partial\Omega} +   
\langle\langle \frac{\partial {\cal L}(\hat\Omega,m^a)}{\partial\Omega} \rangle\rangle 
\label{saddle}
\eeq
with,
\beq
\langle\langle {\cal O}(m) \rangle\rangle = \
\frac{1}{\int \prod_a dm^a \ e^{{\cal L}(\Omega,m^a)}}\ 
\int \prod_a dm^a \  {\cal O}(m^a) \ e^{{\cal L}(\Omega,m^a)}  \ . 
\eeq
The following saddle point equations are easily solved:
\beqa
\frac{\partial\Sigma_0}{\partial W_{ab}}&=&0 \quad\Rightarrow\quad \
W_{ab}=\frac{1}{\beta^2} w^{ab} + r\, q_{ab} \non \\
\frac{\partial\Sigma_0}{\partial T_{ab}}&=&0 \
\quad\Rightarrow\quad T_{ab}=-\frac{1}{\beta^2} t^{ab}  \non \\ 
\frac{\partial\Sigma_0}{\partial L_{ab}}&=&0 \quad\Rightarrow\quad l^{ab}=\frac{\beta^2}{2}\, q_{ab}\ .
\label{partial}
\eeqa
In order to have expressions as similar as possible to previous investigations \cite{bm1,brst1}, 
we define,
\beqa
B_{ab}&=&\beta^2(1-q_{aa}) \delta_{ab} + t^{ab} \non 
\\
- \Delta_{ab}&=&\beta^2(1-q_{aa}) \delta_{ab} + w^{ab} \ ,
\label{defa}
\eeqa
and by using the explicit forms of $\phi_1$, $\phi_2$ we finally obtain,
\beqa
\Sigma_0(\Omega)&=& \
\frac{1}{2\beta^2}\sum_{ab} (B_{ab}^2-\Delta_{ab}^2) - 
\sum_a\left(B_{aa}+\Delta_{aa}\right) (1 - q_{aa}) \non \\
&-& \sum_{ab}\left[
\frac{\beta^2}{4}r^2 \,q_{ab}^2 + \lambda^{ab} q_{ab} - r  \Delta^{ab} q_{ab} \right]
- \frac{1}{2} \log[(2\pi\beta^2)^n \det q_{ab}] 
\label{sigma} 
\\
{\cal L}(\Omega,m^a)&=& - r \sum_a \phi_0(q_{aa},m^a) 
+ \sum_{ab} \lambda^{ab} m^a m^b + \log\det\left(\frac{\delta_{ab}}{1-m_a^2}+B_{ab}\right) 
\non
\\
&-& \frac{1}{2\beta^2}\sum_{ab} 
\left[\tanh^{-1}m^a - \sum_c \Delta^{ac} m^c\right] 
q_{ab}^{-1} 
\left[\tanh^{-1}m^b - \sum_c \Delta^{bc} m^c\right] \ .
\label{elle}
\eeqa
This quantity has to be extremized with respect to the variational parameters
$q_{ab}, B_{ab}, \Delta_{ab}, \lambda_{ab}$. This task is technically very hard
if we assume a nontrivial form for the various matrices. In particular, if we want 
to perform a $k$-RSB calculation of the complexity, at any value of $k$, the situation 
is practically hopeless. What we shall see in the next section is that the BRST supersymmetry 
dramatically simplifies the equations, leaving basically just one matrix, $q_{ab}$, to be fixed
variationally, as in the static calculation.

\subsection{Using the BRST the supersymmetry}

In \cite{juanpe} it was shown that the action in (\ref{action}) 
is invariant under a generalization of the Becchi-Rouet-Stora-Tyutin supersymmetry 
\cite{brs,tito}: if $\epsilon$ is an infinitesimal Grassmann parameter, 
(\ref{action}) is invariant under the transformation,
\beq
\delta m_i = \epsilon\, \psi_i \quad\quad
\delta x_i = \epsilon \, r \, \psi_i \quad\quad
\delta \bar\psi_i = -\epsilon\, x_i \quad\quad
\delta \psi_i = 0 
\ .
\label{trans}
\eeq
The invariance of the action implies that also the average of any function 
of the fields $m, \bar\psi, \psi, x$ must be invariant as well. Thus, if we set to zero  
the variation of $m_i \bar\psi_i$ and  $x_i\bar\psi_i$ \cite{juanpe,brst1},  we obtain 
the two BRST equations,
\beqa
\langle \bar\psi_i \psi_i \rangle &+& \langle m_i x_i \rangle =0 
\label{brst1}
\\
r \,\langle \bar\psi_i \psi_i \rangle &+& \langle x_i x_i \rangle =0  \ \ .
\label{brst2}
\eeqa
Once we replicate the action, from equations (\ref{deltas}) and (\ref{defa}) we have,
\beqa
\langle \bar\psi^a\psi^b\rangle &=& T_{ab} = -B_{ab}/\beta^2 + (1-q_{aa}) \, \delta_{ab} \non \\
\langle x^a m^b\rangle &=& R_{ab} = -\Delta_{ab}/\beta^2 - (1-q_{aa}) \, \delta_{ab} + q\,  
r \ ,\non \\
\langle x^a x^b\rangle &=& L_{ab} \ .
\eeqa
The first BRST equation therefore becomes,
\beq
\Delta_{ab}= - B_{ab} + \beta^2 q_{ab}\, r \ .
\label{brst11}
\eeq
In order to use the second BRST relation we need the expression for $L_{ab}$, which can be obtained by
the saddle point equation $\partial\Sigma_0/\partial q_{ab}=0$, where $\Sigma_0$ is given in (\ref{pezzo1}).
By doing this, and by using the first BRST equation, we obtain:
\beq
\lambda^{ab}= \frac{r}{2} \, \Delta_{ab} \ .
\eeq
As we see, thanks to the BRST relations the variational parameters $\Delta_{ab}$ and $\lambda_{ab}$
can be expressed as functions of $B_{ab}, q_{ab}$ and the parameter $r$. 
From the saddle point equations for the variables $\lambda^{ab}$ and $B_{ab}$ we get the
equations,
\beqa
q_{ab} &=&\langle\langle m^a m^b \rangle\rangle 
\label{sad1} \
\\
B_{ab} &=& - \beta^2 (1-q_{aa}) \delta_{ab}  + 
\beta^2 \langle\langle 
\frac{\partial}{\partial B_{ab}}
\log\det\left( \frac{\delta_{ab}}{1-m_a^2}+B_{ab}\right)
\rangle\rangle 
\label{sad3} 
\eeqa
which are sufficient to fix $B_{ab}$ and $q_{ab}$. It is possible to prove that the remaining
saddle point equations are automatically satisfied by the BRST expression for $\Delta_{ab}$ and 
$\lambda_{ab}$, and therefore do not need to be considered. By using the general formula,
\beq
\frac{\partial \log\det M_{ab}}{\partial M_{ab}}=(M^{-1})_{ab}
\eeq
it is easy to show that equation (\ref{sad3}) admits the solution $B_{ab}=0$. 
As in the annealed case, this is the solution we adopt.
Thus, the two BRST relations we are left with are,
\beq
\Delta_{ab}= \beta^2q_{ab} \, r 
\label{brst111} 
\eeq
\beq
\lambda^{ab}=\frac{1}{2}\beta^2 r^2 q_{ab} \ ,
\label{brst222} 
\eeq
and the only unknown parameter left is the matrix $q_{ab}$.
If we use the two relations (\ref{brst111}) and (\ref{brst222}) 
in equations  (\ref{sigma}) and  (\ref{elle}), and make the 
change of variable $m^a\to h^a=\tanh^{-1}(m^a)$, we obtain a much simpler expression 
for the quenched thermodynamic potential $\Psi(\beta,r)$, which is one of our main results: 
\beq
\beta\Psi(\beta,r) = 
-\log 2  
+ \frac{\beta^2}{4n}\left[r\sum_{ab}^n q_{ab}^2 - \sum_a^n  (1-q_{aa})^2\right] -
\frac{1}{nr} \log\int\prod_a^n dh^a \
\frac{e^{{\cal F}(h^a;q_{ab},r)}}{\left[2 \pi \beta^2 \det q_{ab}\right]^{1/2}}\ \,  \ ,
\label{generale}
\eeq
with,
\beq
{\cal F}(h^a;q_{ab},r)=-\frac{1}{2\beta^2}\sum_{ab}^n h^a q_{ab}^{-1} h^b +r\sum_a^n \log\cosh h^a \ .
\label{effe}
\eeq
In the formula above the limit $n\to 0$ is understood, and the matrix  $q_{ab}$ has still to be fixed 
by the saddle point equation $q_{ab}=\langle\langle m^am^b\rangle\rangle$, with $m^a=\tanh(h^a)$
and where the distribution $p(h^a)=\exp[{\cal F}(h^a)]/(2\pi\beta^2\det q_{ab})$ must now be used to
compute the average $\langle\langle\cdot\rangle\rangle$.

As we have seen in section 1,
the complexity $\Sigma(\beta,f)$ is just the Legendre transform of $\Psi(\beta,r)$. Thus, the quenched 
supersymmetric TAP complexity of the SK model is given by,
\beq
\Sigma(\beta,f) = \beta r f - \beta r \, \Psi(\beta,r)  \ ,
\label{complex}
\eeq
where the parameter $r=r(\beta,f)$ is fixed by the equation,
\beq
\Psi(\beta,r) + r \,\frac{\partial\Psi(\beta,r)}{\partial r}=f \ .
\eeq

\subsection{A special case of the BRST supersymmetry: the Bray-Moore action}

Before proceeding, we discuss here a point which was slightly confusing when the comparison 
was made between the past calculations of the complexity and the most recent ones. 
In all the classic calculations (both annealed and quenched) performed by Bray and Moore \cite{bm3}, 
and also by Bray, Moore and Young
\cite{bmy},  the $J$-dependent part of $F_{TAP}(m)$ in the action was eliminated by using the equations 
$\partial_i F_{TAP}(m)=0$ enforced by the $\delta$-function. More specifically, it was
used the equation,
\beq
-\frac{1}{2} \sum_{ij} J_{ij} m_i m_j=-\frac{1}{2\beta } \sum_{i} m_i \phi_1(q,m_i) \ ,
\label{trick}
\eeq
which is valid in the TAP states.
This substitution simplifies considerably the calculation, but the action obtained in this way
is no longer invariant under (\ref{trans}). 
For this reason, in our former calculation of \cite{brst1}, as in the present one, we used the 
{\it full} form of $F_{TAP}(m)$, equation (\ref{ftap}), and due to this the comparison of our results 
with those of \cite{bm1,bm3,bmy} proved somewhat difficult. 
Moreover, it was not clear what was the equivalent of the BRST complexity, when one used the 'trick' 
(\ref{trick}), which seemed to break the BRST invariance from the outset.

The situation gets much clearer once we realize that the action used by Bray and Moore {\it is}
actually invariant, under a slightly modified version of the BRST supersymmetry. More precisely, 
the Bray-Moore action,
\beq
S_{\rm BM}(m,x,\bar\psi,\psi)= \sum_i x_i \partial_i F_{TAP}(m) + \sum_{ij}\bar\psi_i  \psi_j  
\partial_i \partial_j F_{TAP}(m) - r\, G_{\rm BM}(m) \ ,
\label{bmaction}
\eeq
where,
\beq
G_{\rm BM}(m) = -\frac{1}{2\beta } \sum_{i} m_i \, \phi_1(q,m_i) + \frac{1}{\beta}\sum_i \phi_0(q,m_i)
\ ,
\eeq
is invariant under the following modified BRST transformations,
\beq
\delta m_i = \epsilon\, \psi_i \quad\quad
\delta x_i = \frac{1}{2} \, \epsilon \, r \, \psi_i \quad\quad
\delta \bar\psi_i = -\epsilon\, \left( x_i  + \frac{1}{2} \, r\,  m_i \right)              \quad\quad
\delta \psi_i = 0 
\ ,
\label{bmtrans}
\eeq
to be compared with (\ref{trans}). 
Choosing the same two observables as in the former section, 
we get the modified BRST identities, 
\beqa
\langle \bar\psi_i \psi_i \rangle + \langle m_i x_i \rangle -\frac{1}{2} r\, \langle m_i m_i \rangle &=&0 
\label{bmbrst1}
\\
\frac{1}{2} r \,\langle m_i x_i  \rangle  
+ \frac{1}{2} r \,\langle \bar\psi_i \psi_i \rangle 
+ \langle x_i x_i \rangle &=& 0  \ \ , 
\label{bmbrst2}
\eeqa
(we recall that in order to make a comparison with former calculations we have to set 
$r=-u$). In terms of the variational parameters introduced in the previous sections, 
and once we set $B=0$, we obtain,
\beq
\Delta_{ab}= \frac{1}{2} \beta^2q_{ab} \, r 
\label{bmbrst111} 
\eeq
\beq
\lambda_{ab}=\frac{1}{8} \beta^2 r^2 q_{ab} \ .
\label{bmbrst222} 
\eeq
It is interesting to observe that if we make the change of variable suggested in \cite{leuzzi},
\beqa
\Delta_{ab} &\to& \Delta_{ab} + \frac{1}{2}\, \beta^2 r q_{ab} \\
\lambda_{ab} &\to& \lambda_{ab} + \frac{3}{8}\,  \beta^2 r^2 q_{ab} \ ,  
\eeqa
we can reduce relations (\ref{bmbrst111})-(\ref{bmbrst222}) to 
(\ref{brst111})-(\ref{brst222}), and accordingly reduce the action of Bray-Moore
to our action (\ref{sigma})-(\ref{elle}). This is consistent with the results of
\cite{leuzzi}, and shows that the two calculations performed with or without
the 'trick' (\ref{trick}) are connected by a simple change of variables, under which the 
BRST supersymmetry is conserved.

The important point is that equations (\ref{bmbrst111})-(\ref{bmbrst222}) exactly
coincide with the 'ansatz' used by Bray, Moore
and Young in their calculation of the quenched complexity at the full-RSB level (relations (19) 
of \cite{bmy}), which was consistent with the previous Bray-Moore quenched calculation of 
\cite{bm3}. This fact proves that the {\it quenched} complexity considered in \cite{bm3,bmy} 
was in fact BRST symmetric. On the other hand, as it was shown in \cite{brst1}, 
the {\it annealed} complexity of the total number of TAP states considered by Bray and Moore
in \cite{bm1} was not BRST symmetric. For an extensive discussion of the comparison between the
annealed BRST complexity of \cite{brst1} and the BRST-breaking one of \cite{bm1}, see \cite{leuzzi}
and \cite{comment}.

\section{Connections with the static free energy}

Now that we have obtained a general quenched expression for the thermodynamic potential 
$\Psi(\beta,r)$, we want to investigate what are the connections with 
the standard thermodynamic potential, that is the free energy of the system $F(\beta)$.
We recall that the thermodynamic potential $\Psi(\beta,r)$, is the constrained free energy 
density (per replica) of $r$ real replicas forced to stay in the same metastable state, and thus
by definition we have $F(\beta)=\Psi(\beta,r=1)$. This identity was first proved by De Dominicis 
and Young in \cite{ddy}, with the assumptions of a particular ansatz, which was shown in \cite{brst1}
to be nothing else than the BRST relations. Our goal now is to investigate further the
relation between $\Psi$ and $F$ at a generic value of $r\neq 1$.

\subsection{Complexity vs statics: a preliminary step}

First of all, we note that the {\it annealed} calculation of $\Psi(\beta,r)$
is equivalent to assume for the TAP overlap 
matrix in equation (\ref{generale}) the simple form $q_{ab}= q\,\delta_{ab}$. If this is done, 
it is straightforward to prove (as it was done in \cite{brst1}) that the annealed potential $\Psi$ is 
equal to the quenched static free energy, calculate at the one step of replica symmetry breaking, with 
self overlap $q_1=q$, mutual overlap $q_0=0$ and replica symmetry breaking point $x=r$. In the
annealed case we obviously have just one value of the overlap $q$, and therefore in \cite{brst1} 
we could only find a connection with $F_{1RSB}$, once we set $q_0=0$. Now that we have done the quenched
calculation it is natural to expect that the $0$RSB potential $\Psi_{0RSB}$ is connected to the $1$RSB 
static free energy $F_{1RSB}$, with mutual overlap $q_0\neq 0$. Thus, before considering the case of a generic number 
of steps of replica symmetry breaking, we will focus on this simpler case. We assume a $0$RSB form of the TAP overlap 
matrix $q_{ab}$ in (\ref{generale}), that is,
\beq
q_{ab}= q_0 + (q_1-q_0) \, \delta_{ab} \quad\quad\quad [{\rm 0RSB}] \ .
\eeq
The annealed case is recovered by setting $q_0=0$. By using this ansatz in equation (\ref{generale}), 
we find
\beq
\beta\Psi(\beta,r) =  
-\log 2 + \frac{\beta^2}{4}
[(r-1)q_1^2+2q_1-1-rq_0^2]  - \frac{1}{r} I(q_{ab},r)
\label{0rsb}
\eeq
where we have defined,
\beq
I(q_{ab},r)= -\frac{q_0}{2(q_1-q_0)}+ 
\frac{1}{n}\log\int\prod_a^n dh^a \frac{e^{{\cal F}(h^a;q_{ab},r)}}{\sqrt{2\pi\beta^2(q_1-q_0)}} \  \ ,
\label{apple}
\eeq
and the limit $n\to 0$ in the expression for $\det q_{ab}$ has already been taken. In order to proceed 
we note that,
\beq
[q^{-1}]_{ab}= - q_0/(q_1-q_0)^2 + \delta_{ab}/(q_1-q_0) 
\eeq
The term $\sum_{ab}h^a q_{ab}^{-1} h^b$ in the exponential in $\cal F$ can be thus rewritten as,
\beq
\sum_{ab}^nh^a q_{ab}^{-1} h^b = 
\frac{1}{q_1-q_0}\sum_a^n h_a^2 
- \frac{q_0}{(q_1-q_0)^2} \left(\sum_a^n h^a\right)^2 \ .
\eeq
By using the following Hubbard-Stratonovich identity,
\beq
\exp\left[\frac{q_0}{2 \beta^2(q_1-q_0)^2}\left(\sum_a h^a\right)^2\right]
=\int\frac{dz}{\sqrt{2 \pi q_0}}\ 
\exp\left[-\frac{z^2}{2 q_0}+\frac{z}{\beta(q_1-q_0)}\sum_a h^a \right] \ ,
\eeq
we find that replicas factorizes in the integral in $I(q_{ab},u)$ and thus we can pass from the
$n$ variables $h_a$, to one single scalar variable $h$. After making the further change of variable 
$h=\beta(z+y)$, we finally obtain,
\beq
I(q_{ab},r) = 
\int\frac{dz}{\sqrt{2 \pi q_0}}e^{-\frac{z^2}{2 q_0}}\  
\log\int\frac{dy}{\sqrt{2 \pi (q_1-q_0)}}\ e^{- \frac{y^2}{2 (q_1-q_0)}} \,
\cosh^r\left[\beta(z+y)\right]
\ .
\eeq
If we substitute this form into equation (\ref{0rsb}), we find that we have exactly reconstructed the 
static free energy of the SK model at the $1$RSB level of approximation \cite{1rsb,mpv}, where the replica 
symmetry breaking point (normally called $x$) is equal to the parameter $r$. Therefore we have,
\beq
\Psi_{0RSB}(\beta,r) = F_{1RSB}(\beta) \ .
\eeq
We stress that the overlap matrix $Q_{\alpha\beta}$ of the static free energy has a different structure from the overlap 
matrix $q_{ab}$ of the potential $\Psi$. More precisely, for $\Psi_{0RSB}$ we have the two parameters overlap 
matrix,
\beq
q_{ab}= q_0 + (q_1-q_0)\, \delta_{ab} \ .
\eeq
Moreover, $\Psi_{0RSB}$ is a function of the variable $r$. On the other hand, for the static free energy 
$F_{1RSB}$ we have the three parameters overlap matrix $Q_{\alpha\beta}$,
\beq
Q_{\alpha\beta}= q_0 + (q_1-q_0)\, \varepsilon_{\alpha\beta}^{(r)} + (1-q_1)\, \delta_{\alpha\beta} \ ,
\eeq
where $\varepsilon_{\alpha\beta}^{(r)}$ is a block ultrametric matrix, equal to $1$ within a diagonal block of
size $r$ and zero elsewhere. In other words, the variable $r$ of the potential $\Psi$ calculated at the $0$RSB level
becomes the replica symmetry breaking point of the static free energy $F$ calculated at the $1$RSB level.

This result is 
in accordance with what found in other systems solved exactly by one step of replica symmetry breaking, 
where the $0$RSB complexity is always related to the $1$RSB static free energy by means of a 
Legendre transform with respect of the breaking point $x$ \cite{monasson}.  
In the SK model, however, the statics is solved by a full replica symmetry breaking ansatz, with an infinite
number of breaking points. It was therefore not clear whether the Legendre relation between complexity and 
free energy was preserved in the SK model. More practically,
once we consider $k$RSB solutions, we have more than one breaking parameter $x$, and thus it is 
not obvious which one of the $k$ breaking points must be used to perform the Legendre transform. 
What we show in the next section is that the constrained thermodynamic potential $\Psi(\beta,r)$
computed at the $k$ RSB level is identical to the static free energy $F(\beta)$ calculated at the
$k+1$ RSB level, with $r$ being equal to the {\it largest} breaking point $x_{max}$. This means that 
the Legendre relation between TAP complexity and static free energy is conserved at any level of replica symmetry 
breaking, and that the Legendre parameter is the largest replica symmetry breaking point.

\subsection{Complexity vs statics: the general case}

We start by recalling the general from of the static quenched free energy in the SK model,
before any ansatz on the overlap matrix is done \cite{sk},
\beq
\beta F= - \frac{\beta^2}{4}+ \frac{\beta^2}{2n_s}\sum_{\alpha>\beta}^{n_s} Q_{\alpha\beta}^2 
- \frac{1}{n_s}\log\sum_{[\sigma^\alpha]}\exp\left[\frac{\beta^2}{2}
\sum_{\alpha\neq \beta}^{n_s} Q_{\alpha\beta}\sigma^\alpha \sigma^\beta \right]\ ,
\label{statica}
\eeq
where $Q_{\alpha\beta}$ is a $n_s\times n_s$ matrix, with $n_s\to 0$. The subscript in $n_s$ stands for {\it static},
and it is necessary in order to distinguish the size of the static overlap matrix $Q_{\alpha\beta}$ from the size $n$ of 
the TAP overlap matrix $q_{ab}$.
As a first step to make the general expression (\ref{generale}) for $\Psi(\beta,r)$ a bit closer 
to the static free energy (\ref{statica}), we can write,
\beq
\cosh(h_a)^{r} = \frac{1}{2^r}\, \sum_{[\tau_a^\mu=\pm 1]}e^{h_a\, \sum_\mu^r \tau_a^\mu}
\eeq
such that in the expression of $\Psi$  we can integrate now over the Gaussian variables $h_a$ and obtain,
\beq
\beta\Psi(\beta,r)
= -\frac{\beta^2}{4}  +\frac{\beta^2}{4n} (r-1)\sum_a^n q_{aa}^2 +
\frac{\beta^2}{2n} r \sum_{a>b}^n q_{ab}^2
+ \frac{\beta^2}{2n}\sum_a^n q_{aa}-\frac{1}{nr} \log\sum_{[\tau_a^\mu]}
\exp\left[\frac{\beta^2}{2}\sum_{ab}^n \sum_{\mu \nu}^r \tau_a^\mu q_{ab} \tau_b^\nu\right]
\eeq
We want to prove that $\Psi(\beta,r) = F(\beta)$ if some suitable structures for the overlap 
matrices are considered. It is straightforward 
to see that this equation is fulfilled if the two following relations
are satisfied,
\beqa
\frac{1}{n r} \log\sum_{[\tau_a^\mu]} 
\exp\left[\frac{\beta^2}{2}\left(
\sum_{ab}^n\sum_{\mu\nu}^r \tau_a^\mu q_{ab} \tau_b^\nu - \sum_a^n\sum_\mu^r q_{aa}
\right)\right] 
&=& 
\frac{1}{n_s} \log\sum_{[\sigma^\alpha]}\exp\left[\frac{\beta^2}{2}\sum_{\alpha\neq \beta}^{n_s} 
Q_{\alpha\beta}\sigma^\alpha \sigma^\beta \right] 
\label{pippa}
\\
\frac{r-1}{2 n} \sum_a^n q_{aa}^2 + \frac{r}{n}\sum_{a>b}^n q_{ab}^2
&=&
\frac{1}{n_s}\sum_{\alpha>\beta}^{n_s} Q_{\alpha\beta}^2 
\label{pappa}
\eeqa
The first thing to note is that the form of the equations suggests the identity,
\beq
n_s= r \cdot n \ .
\eeq
Once this identification is done, we can connect the $\sigma^\alpha$ spin variables ($\alpha=1,\dots, n_s$), to the
$\tau_a^\mu$ spin variables ($a=1,\dots,n;\  \mu=1,\dots, r$) in the following way,
\beq
\left(\sigma_1,\dots ,\sigma_{n_s}\right)
= \left(\tau_1^1,\dots ,\tau_1^r,\tau_2^1,\dots ,\tau_2^r, \ \dots \dots\dots\  , \tau_n^1,\dots ,\tau_n^r\right)
\label{stringa}
\eeq
Let us now assume that we are performing the calculation of the potential $\Psi(\beta,r)$ at $k$ RSB level. 
In this case the TAP overlap matrix $q_{ab}^{(k)}$ is given by,
\beq
q_{ab}^{(k)} =q_0 +  \sum_{i=1}^{k+1}(q_i-q_{i-1}) \ \varepsilon_{ab}^{(n, y_{i})}  \ , 
\eeq
with,
\beq
y_{k+1}=1 \ ,
\quad\quad  \varepsilon_{ab}^{(n,1)}=\delta_{ab} \ .
\eeq
The matrices $\varepsilon^{(n, y_{i})}$ are the $n\times n$ ultrametric block matrices, equal to
one on the diagonal blocks of size $y_i$ and zero elsewhere.
The variables $y_i$ are thus the replica symmetry breaking points. Unlike the standard static case, 
in the TAP approach the diagonal of the overlap matrix $ q_{aa}= q_{k+1}$ is essential, as it contains  
the self-overlap of the TAP states. For this reason the largest breaking point is trivial, that is
$y_{k+1}=1$. Therefore, there are $k+1$ values of the overlap, but only $k$ nontrivial breaking points, 
ans thus $q_{ab}$ is indeed a $k$RSB matrix. The simplest example of this matrix was shown in the last section.

By using relation (\ref{stringa}) it is possible to prove the following key formula,
\beq
\sum_{ab}^n\sum_{\mu\nu}^r \varepsilon_{ab}^{(n,y_i)}\, \tau_a^\mu \tau_b^\nu 
=
\sum_{\alpha\beta}^{rn} \varepsilon_{\alpha\beta}^{(rn,ry_i)}\, \sigma_\alpha \sigma_\beta \ ,
\eeq
and we recall that $rn=n_s$. It is now possible to prove that,
\beq
\sum_{ab}^n\sum_{\mu\nu}^r \tau_a^\mu q_{ab}^{(k)} \tau_b^\nu - \sum_a^n\sum_\mu^r q_{aa}^{(k)}
\ = \ 
\sum_{\alpha\neq \beta}^{rn} Q_{\alpha\beta}^{(k+1)}\, \sigma_\alpha \sigma_\beta\, \ ,
\label{stella}
\eeq
with,
\beq
Q^{(k+1)}_{\alpha\beta}
= q_0 + \sum_{i=1}^{k+1}(q_i-q_{i-1}) \ \varepsilon_{\alpha\beta}^{(rn, r y_{i})} \ .
\label{ququ}
\eeq
Equation (\ref{pippa}) is therefore verified by the $rn\times rn$ ultrametric matrix 
$Q_{\alpha\beta}^{(k+1)}$, which has a standard RSB form with $k+1$ levels of replica symmetry breaking.
The entries of $Q_{\alpha\beta}^{(k+1)}$ are the same as the TAP overlap matrix $q_{ab}^{(k)}$, 
with the only important difference that the elements on the diagonal of $Q_{\alpha\beta}^{(k+1)}$
are irrelevant, since the sums in (\ref{statica}) are only performed for $\alpha\neq \beta$.
On the other hand, it is clear from (\ref{ququ}) that 
the $k+1$ replica symmetry breaking points $(x_1,\dots , x_{k+1})$ of the
matrix $Q_{\alpha\beta}^{(k+1)}$ are given by,
\beqa
x_1&=&r\,y_1 \non
\\
&\dots& \non 
\\
x_k&=&r \,y_k \non
\\
x_{k+1}&=&r\, y_{k+1}= r
\ .
\eeqa
In other words, the parameter $r$ represents the {\it largest} breaking point of the static overlap 
matrix $Q_{\alpha\beta}^{(k+1)}$.

Finally, from the simple relation,
\beq
\frac{r}{n}\sum_{a\neq b}^n \varepsilon_{ab}^{(n,y_i)} = 
\frac{1}{rn}\sum_{\alpha\neq \beta}^{rn} \varepsilon_{\alpha\beta}^{(rn,ry_i)} 
\eeq
it is straightforward to check that also relation (\ref{pappa}) is verified.

Summarizing, what we have proved is that the thermodynamic potential $\Psi(\beta,r)$ 
calculated at the $k$ RSB level is equal to static free energy $F(\beta)$ calculated at 
the $k+1$ RSB level. In other words, the free energy of $r$ real replicas forced to be in 
the same state, computed at the $k$ RSB level, is equal to the ordinary free energy of one 
single system, computed at the $k+1$ RSB level, where the extra $k+1$-th symmetry breaking 
point is equal to $r$. The symmetry breaking points of the static matrix $Q_{\alpha\beta}^{(k+1)}$ 
are simply the ones of the TAP matrix $q_{ab}^{(k)}$ rescaled by the parameter $r$. 
We can write our result as,
\beq
\Psi\left(\beta,r; q_{ab}^{(k)}\right) = F\left(\beta; Q_{ab}^{(k+1)}\right) \ ,
\eeq
where the relation between $q_{ab}^{(k)}$ and $Q_{ab}^{(k+1)}$ is given above, and,
as we have seen, the parameter $r$ is the largest replica symmetry breaking point 
of the static matrix $Q_{ab}^{(k+1)}$.
From equation (\ref{complex}), and given the relation between $\Psi(\beta,r)$
and $F(\beta)$, we finally have the general Legendre equation connecting the supersymmetric 
complexity of the TAP states to the standard static free energy in the SK model,
\beq
\Sigma(\beta,f) = \beta x f - \beta x F(\beta; x) \ ,
\label{tuka}
\eeq
with $x$ fixed by the Legendre equation,
\beq
f = F(\beta; x) + x\, \frac{\partial F(\beta; x)}{\partial x} \ .
\label{stuka}
\eeq
This result can be summarized by saying that {\it the quenched complexity of the TAP states is
the Legendre transform of the static free energy with respect to the largest breaking point $r$ 
of its overlap matrix}. Note that from equation (\ref{tuka}) it is trivial to check that consistency 
with the statics is obtained at {\it any} level of RSB. Indeed, $\Sigma=0$ for $f=F$, i.e. the lower
band edge coincides with the static free energy; moreover, the derivative of $\Sigma$ with respect to
the free energy $f$ at the lower band edge is just $\beta x$, where $x$ is the largest static replica
symmetry breaking point, as it should be. The fact that this consistency can be so transparently read 
from equation (\ref{tuka}) is one of the  main
virtues of the Legendre-supersymmetric approach.

If we invert our point of view, and fix the largest breaking parameter of the static overlap matrix 
$Q_{\alpha\beta}$ to its equilibrium value $x_{max}$, and let $r$ free, we can give a deeper interpretation 
of these results. First, let us recall that the symmetry breaking points of the 
overlap matrix are related to the overlap probability distribution $P(q)$ \cite{mpv}. In 
particular, the probability of the TAP self-overlap $q_{\alpha\alpha}$, is given by,
\beq
w_{\alpha\alpha}= 1-y_{max} \ .
\eeq
Therefore, if we have,
\beq
r \geq x_{max} \quad\Rightarrow\quad y_{max}= \frac{x_{max}}{r} \leq 1 \ ,
\eeq
the probability of the TAP self-overlap, $w_{\alpha\alpha}$ is different from zero, 
and thus the number of TAP states is {\it not} exponentially large in $N$. This is consistent with the general 
philosophy of the Legendre transform method \cite{monasson}, which states that if the parameter $r$ is larger than the static 
equilibrium breaking point $x_{max}$, we are stuck at the equilibrium ground states, and therefore 
$f=F(\beta)$ and from equation (\ref{stuka}) we obviously find $\Sigma=0$. 
Thus, the complexity of the static equilibrium 
states is by definition zero. If, on the other hand, 
\beq
r < x_{max} \quad\Rightarrow\quad y_{max}= \frac{x_{max}}{r} > 1 \ ,
\eeq
then the weight of the TAP self-overlap is zero. In this phase of $r$ the number of TAP states is exponentially
large in $N$, their complexity is nonzero and their free energy density $f$ is larger than the equilibrium 
value $F(\beta)$.
If we parametrize the ultrametric matrices  $q_{ab}$ and $Q_{\alpha\beta}$ by means of the two functions
$q(y)$ and $Q(x)$ respectively, we can write,
\beq
Q(r\cdot y)=q(y) \ .
\eeq
This last relation explains quite well the fact that the standard static 
approach and the TAP approach are simply connected by a coarse graining relation,
where $r$ is the scale parameter: what we are doing by using the TAP approach is to
start at a higher level in the coarse graining procedure, since the elementary objects
are {\it states}, rather than {\it configurations} as in the static approach.
We would like to remark that the many of the physical considerations of 
this last paragraph were already 
pointed out in \cite{bmy}, where, among the other things, it was for the 
first time uncovered the Legendre transform relationship between complexity and 
free energy, and the role of $r$ as an exploring tool of the spectrum of states.
However, the fact that the modified BRST symmetry (\ref{bmtrans}) of that calculation was
not recognized, prevented 
to prove equation (\ref{tuka}), and the consistency between TAP complexity and 
static results was quantitatively tested only close to the lower band edge 
and to the critical temperature. In our case, from (\ref{tuka}) we
automatically have this consistency at any value
of $f$ and $T$.

A final  point to note is that if we set $r=1$ the two matrices $q_{ab}^{(k)}$ and 
$Q_{\alpha\beta}^{(k+1)}$ are actually the same, since the largest breaking point of 
$Q_{\alpha\beta}^{(k+1)}$ is one, and thus this is in fact a $k$RSB matrix.  This is 
consistent with the fact that for $r=1$ the 
thermodynamic potential $\Psi(\beta,r)$ is just the standard static free energy \cite{ddy}

\section{conclusions}

The main result of this work is given by equations (\ref{tuka}) and (\ref{stuka}): 
the BRST quenched complexity of the TAP states in the SK model is just the Legendre transform 
of the static free energy. The key tool for obtaining this result has been the  
supersymmetry. Our result can be reinterpreted by saying that the degree of difficulty
of the computation of the complexity becomes equal to the one of the standard free energy
once the BRST symmetry is used. This is an important point. In principle, from a technical 
point of view these calculations are quite different. For the complexity we need
the TAP free energy $F_{TAP}$, its derivative and its Hessian, while for the static partition 
function we simply need the Hamiltonian $H$. Even though the relationship between $F_{TAP}$ 
and $H$ may be easy to uncover \cite{tap}, it still would seem that the calculation of the
complexity has a higher degree of difficulty, since it also involves $\partial F_{TAP}$,
and $\partial^2 F_{TAP}$. This is exactly the technical problem that the  old calculations
of the TAP complexity had to face, whenever a connection with the statics was investigated.
These calculations treated $F_{TAP}, \partial F_{TAP}$ and $\partial^2 F_{TAP}$ as
independent functions, while clearly they are not. In so doing an important
physical information was wasted.

The BRST supersymmetry exactly takes care of the fact that $F_{TAP}, \partial F_{TAP}$ and 
$\partial^2 F_{TAP}$ are {\it not} independent, and therefore it reduces the 
redundant difficulty of the calculation, making it practically equivalent to the one 
of the standard free energy. This fact suggests that the BRST supersymmetry should be 
considered an essential tool each time the complexity of a glassy system has to be computed, 
since it encodes at the deepest level the natural connections between the state function (which in
structural glasses may be the Hamiltonian) and 
its topological properties, expressed in the distribution of the metastable states.
However, we know that in general there may be some solutions of the saddle points
equations which {\it break} the BRST symmetry, still giving a nontrivial complexity.
At the moment, it is unclear what is the physical meaning of these non-BRST saddle points, nor
whether they should be preferred to the BRST ones \cite{comment,leuzzi}.
What we believe it can be said with some confidence, is that the consistency with the
statics, i.e. the equalities between static ground state and lower band edge, and
between breaking parameter $x$ and $\Sigma'(f)$, {\it is} given by the BRST
saddle point. Indeed, as already stressed, the present method is only valid in the energy phase
where stable minima are dominant over unstable saddles, and this is certainly true
at, or very close to, the lower band edge. Whether the BRST solution of the saddle point
equations is the only relevant one also at higher energies, depends on the mutual 
distribution of minima and saddles. Of course, the same uncertainty also holds for any 
other non-BRST solution.

As we have seen the complexity of the TAP {\it states} is given by the Legendre transform of
the free energy with respect to the {\it largest} breaking point of the overlap matrix. It
has been recently conjectured in \cite{riko} that when there is more than one step 
of replica symmetry breaking the complexity of the {\it clusters} at level $i$ may be 
given by the Legendre transform with respect to the breaking point $x_i$. This seems 
a sound generalization of our result, since the deepest clusters, associated to the 
largest breaking point, are indeed the states. Moreover, this conjecture
raises the interesting issue whether, in some cases, it may be more relevant from a 
dynamical point of view to calculate the complexity of the clusters, rather than the one 
of the states. It would be interesting to study whether the conjecture of \cite{riko} can
be exactly proved, perhaps using the supersymmetric approach of the present work.

We thank Alan Bray, Luca Leuzzi, Mike Moore, and Tommaso Rizzo for many stimulating
discussions.

\end{document}